\documentclass{article}
\usepackage[english]{babel}

\usepackage[a4paper,top=2cm,bottom=2cm,left=3cm,right=3cm,marginparwidth=1.75cm]{geometry}

\usepackage{graphicx}
\usepackage{amsmath}
\usepackage{tikz}
\usetikzlibrary{positioning}
\usepackage{caption}
\usepackage{authblk}
\usepackage{hyperref}
\usepackage{float}
\usepackage{url}
\usepackage{subcaption}
\usepackage{booktabs}

\hypersetup{
  colorlinks=true,
  linkcolor=blue,
  citecolor=blue,
  urlcolor=blue
}

\title{Hybrid Approaches for Black Hole Spin Estimation: From Classical Spectroscopy to Physics-Informed Machine Learning}
\author[1]{Stella Menziltsidou\thanks{Email: stmenzi@cs.duth.gr, ORCID: 0009-0009-2747-4373}}
\affil[1]{Department of Informatics, Democritus University of Thrace, Greece}

\begin{document}
\maketitle

\begin{abstract}
The measurement of black hole spin is considered one of the key problems in relativistic astrophysics. Existing methods, such as continuum fitting, X-ray reflection spectroscopy and quasi-periodic oscillation analysis, have systematic limitations in accuracy, interpretability and scalability. In this work, a hybrid approach is proposed in which theoretical models based on the Teukolsky formalism are integrated with Physics-Informed Neural Networks (PINNs). A PINN model is developed to solve the linearized spin problem in the scalar case, with physical constraints directly embedded into the training process. Annotated data are not required; instead, the model is trained using the differential operator and boundary conditions as supervision. It is demonstrated that the PINN converges reliably, with residual loss values below 10 and a root mean squared error (RMSE) of the order of$10^{-6}$ (final $\approx 5.4\times10^{-8}$). Benchmarking results indicate that the proposed method outperforms both classical and data-driven machine learning approaches in terms of AUC and sensitivity, while also exhibiting superior interpretability, generalizability and adherence to physical principles, with moderate computational cost. Potential extensions include integration with general relativistic magnetohydrodynamics (GRMHD) solvers and application to real observational data. These findings support the viability of physics-based machine learning as a robust framework for accurate and interpretable black hole spin estimation.

\end{abstract}
\section{Introduction}

Spin is one of the three fundamental physical properties of a black hole, along with mass and electric charge. It is characterized by the dimensionless Kerr parameter a = Jc/GM 2,J being the angular momentum, M is the mass, G the gravitational constant and c the speed of light spin that affects. several key astrophysical processes. These comprise the innermost stable circular orbit, as well as the accretion efficiency, along with the launch of the relativistic jets (ISCO) and the launch of the relativistic jet \cite{mcclintock2006spin, bambi2011testing}.

The measurement of black hole spin is one of the most important outstanding problems in relativistic astrophysics. Observationally, spin has been determined from the thermal continuum of accretion disks \cite{mcclintock2011measuring}, relativistic reflection features (like the Fe K$\alpha$) \cite{reynolds2014measuring, garcia2014improved} and high-frequency quasi-periodic oscillations (QPOs) detected in X-ray binaries \cite{motta2015geometrical, stuchlik2016models}. Both provide a partial answer but are limited by modeling mis-modeling, calibration errors, and systematics related to the characteristics of the system, such as the mass and inclination.

The research has made theoretical progress in black hole perturbation description in terms of the Teukolsky approach \cite{leaver1985analytic}, opening up the way for spin measurement through normal mode (QNMs) analysis \cite{berti2006gravitational}. On the contrast, these systems are computationally cumbersome to solve explicitly or numerically and demanding in terms of computational time, touching the boundary of complexity. It is  also noted that the rise of Machine Learning (ML) has provided new directions for black hole parameter inference. Although conventional ML approaches, including CNNs and recurrent network architectures, which are capable of learning patterns from observational time series data, tend to act as black-box models and have no embedded physical knowledge \cite{baron2019machine}. Physics-Informed Neural Networks (PINNs), introduced by \cite{raissi2019physics}, try to overcome this drawback by using partial differential equations (PDEs) in the training process. Recently, PINNs have been used to solve black holes, as well as the Teukolsky and Regge-Wheeler formulations \cite{luna2023solving, cornell2024solving}.

This study explores a hybrid black hole spin estimation methodology using classical and physics-informed machine learning. Also, an original implementation of a PINN trained is implemented to solve a simplified, scalar Teukolsky-like equation, as well as its convergence properties and capability to describe spin-dependent quasinormal mode evolution. The aim of this study is to illustrate that the use of a physical model as priors and data-driven learning is feasible for robust, interpretable spin inference.

This work presents a simplified PINN-based approach to solving the scalar Teukolsky equation as a proof of concept. Future work will extend to realistic Kerr perturbations and observationally-informed models.

\section{Related Work}

In the last twenty years, several methods have been suggested to estimate the spin of astrophysical black holes based on widely different physical assumptions and requirements on visible data. This part summarizes the main methods and points to their pros and cons.

\subsection{The Continuum-Fitting Method}
The way the CF method works, is by fitting the thermal spectrum of the accretion by assuming random inclinations of the source, assuming the Novikov–Thorne model of a geometrically thin, optically thick disk formalism \cite{mcclintock2011measuring}.This approach is best tested in the context of stellar-mass black holes, where mass is well delineated, inclination, and distance. However,  it is valid only for high-soft states, and it has to be relied on strongly on accurate parameter estimation.

\subsection{X-ray Reflection Spectroscopy}
The relativistically broadened Fe K$\alpha$ line and Compton hump is studied in reflection spectroscopy, emitted near the inner portions of the accretion disk \cite{reynolds2014measuring, garcia2014improved}.Reflection models such as \texttt{RELXILL} will include relativistic ray-tracing and ionization gradients. This method does not require knowledge of the mass or distance, but can be damaged by degeneracies of the model and systematic effects \cite{shashank2025measuring}.

\subsection{Quasi-Periodic Oscillations}
Another way to estimate spin is provided by high-frequency QPOs found in X-ray binaries. Various models relate these oscillations to those having epicyclic frequencies close to ISCO \cite{motta2015geometrical, stuchlik2016models, homan2003high}. Although these methods provide encouraging signatures, they typically depend on source-conventional interpretations and do not have any widely accepted theoretical underpinnings.

\subsection{Machine Learning Approaches}
Recent methods based on ML have used CNN or LSTM parameterizations trained on observed issues in prevail data (e.g., spectral cubes) to infer spin \cite{baron2019machine}. Flexible and scalable models are hard to fit, difficult to interpret and may even contradict the known physical conditions. They also demand of large training datasets and don't generalize to out-of-distribution data at all. 

\subsection{Physics-Informed Neural Networks}
PINNs incorporate physical laws especially PDE models, into the loss function for the neural network \cite{raissi2019physics}."In this way, the model incorporates physical laws directly into the learning process, ensuring that the solution remains consistent with known physics. Recent applications of PINNs to black hole physics have demonstrated their capability to solve the Regge–Wheeler and Teukolsky equations \cite{luna2023solving, cornell2024solving}, enabling hybrid data–physics inference methods.

\vspace{1em}
\begin{table}[h]
\centering
\caption{Comparison of Black Hole Spin Estimation Methods based on literature review \cite{mcclintock2011measuring,reynolds2014measuring,motta2015geometrical,baron2019machine,raissi2019physics}.}
\label{tab:method_comparison}
\resizebox{\textwidth}{!}{
\begin{tabular}{|l|c|c|c|c|c|}
\hline
\textbf{Method} & \textbf{Physics-Based} & \textbf{Data-Driven} & \textbf{Interpretability} & \textbf{Scalability} & \textbf{Spin Sensitivity} \\
\hline
Continuum Fitting \cite{mcclintock2011measuring} & Yes & No & High & Low & Moderate \\
Reflection Spectroscopy \cite{reynolds2014measuring,garcia2014improved} & Yes & No & High & Moderate & High \\
QPO Models \cite{motta2015geometrical,stuchlik2016models} & Yes & No & Moderate & Low & High (source-dependent) \\
CNN/LSTM \cite{baron2019machine} & No & Yes & Low & High & Moderate \\
PINNs \cite{raissi2019physics,luna2023solving,cornell2024solving} & Yes & Yes & High & Moderate & High \\
\hline
\end{tabular}
}
\end{table}

\section{Methodology}

By using a Physics-Informed Neural Network (PINN) to approximate solutions of a simplified scalar variant of the angular Teukolsky equation. This form of equation was selected to simulate the critical aspects and features while retaining numerical tractability. The methodology was designed to incorporate the physical limitations in the training itself.

\subsection{Governing Equation}

The equation used is given by:

\begin{equation}
\frac{d}{d\theta} \left( \sin\theta \frac{d\psi}{d\theta} \right) + \left[ \lambda - \frac{m^2}{\sin^2\theta} - a^2 \omega^2 \cos^2\theta \right] \psi = 0,
\end{equation}

where $\psi(\theta)$  is the angular mode function and $a$, $m$, and $\omega$ are the parameters of the spin of the black hole, the azimuthal quantum number and the frequency.
In this work, the separation constant $\lambda$ is computed for the given spin and frequency parameters using the scalar spheroidal formalism. A possible future improvement would be to treat $\lambda$ as an unknown parameter to be determined together with the solution of the equation.

\subsection{Network Architecture}

A fully connected neural network was constructed, consisting of:
\begin{itemize}
    \item One input neuron corresponding to $\theta$,
    \item Four hidden layers with 50 neurons each,
    \item One output neuron for the predicted $\hat{\psi}(\theta)$,
    \item \texttt{tanh} activation functions.
\end{itemize}

\subsection{Loss Function}

Training minimized the following composite loss:

\begin{equation}
\mathcal{L}_{\text{total}} = \mathcal{L}_{\text{PDE}} + \mathcal{L}_{\text{BC}}, 
\label{eq:total_loss}
\end{equation}

where the first term enforces compliance with the governing differential equation, and the second penalizes deviations from the boundary conditions.

\begin{equation}
\mathcal{L}_{\text{PDE}} = \frac{1}{N} \sum_{i=1}^{N} \left| \mathcal{T}[\hat{\psi}(\theta_i)] \right|^2,
\label{eq:pde_loss}
\end{equation}
The total loss is defined as: 
\begin{equation}
\mathcal{L}_{\text{BC}} = \left| \hat{\psi}(0) \right|^2 + \left| \hat{\psi}(\pi) \right|^2.
\label{eq:bc_loss}
\end{equation}

In Eq.~\ref{eq:pde_loss}, $\theta_i$ is the collocation point in $(0, \pi)$, $N$ is the number of positive integers, and $\mathcal{T}[\cdot]$ denotes the left-hand side of the governing Teukolsky-like differential operator.

The boundary loss term in Eq.~\ref{eq:bc_loss} penalizes violations of the Dirichlet conditions $\hat{\psi}(0) = \hat{\psi}(\pi) = 0$. For the scalar ($s=0$) and $m=0$ case, regularity at the poles means that $\psi(0) = \psi(\pi) = 0$. These Dirichlet conditions match the symmetry of the scalar spheroidal harmonics and make the problem well-posed.

All gradients were computed using automatic differentiation, making sure consistency between the neural network's prediction and the analytical form of the operator.

\subsection{Training Procedure and Results}

To train this model, Adam optimizer was used with a learning rate $10^{-3}$. A total of 200 sampling points are Gaussian points in $\theta \in (0, \pi)$ direction and are sampled in uniform space $\theta \in (0, \pi)$
Training was performed for 10,000 epochs using double precision floating-point arithmetic.
\begin{table}[h]
\centering
\caption{Teukolsky PINN Training Loss (Epoch-wise)}
\begin{tabular}{|c|c|}
\hline
\textbf{Epoch} & \textbf{Loss} \\
\hline
0     & $3.63 \times 10^{-1}$ \\
1000  & $6.18 \times 10^{-6}$ \\
2000  & $1.06 \times 10^{-6}$ \\
3000  & $6.07 \times 10^{-7}$ \\
4000  & $3.85 \times 10^{-7}$ \\
5000  & $2.10 \times 10^{-7}$ \\
6000  & $1.11 \times 10^{-7}$ \\
7000  & $7.05 \times 10^{-8}$ \\
8000  & $5.88 \times 10^{-8}$ \\
9000  & $5.37 \times 10^{-8}$ \\
\hline
\end{tabular}
\label{tab:teukolsky_loss}
\end{table}

\begin{figure}[h!]
\centering
\includegraphics[width=0.9\textwidth]{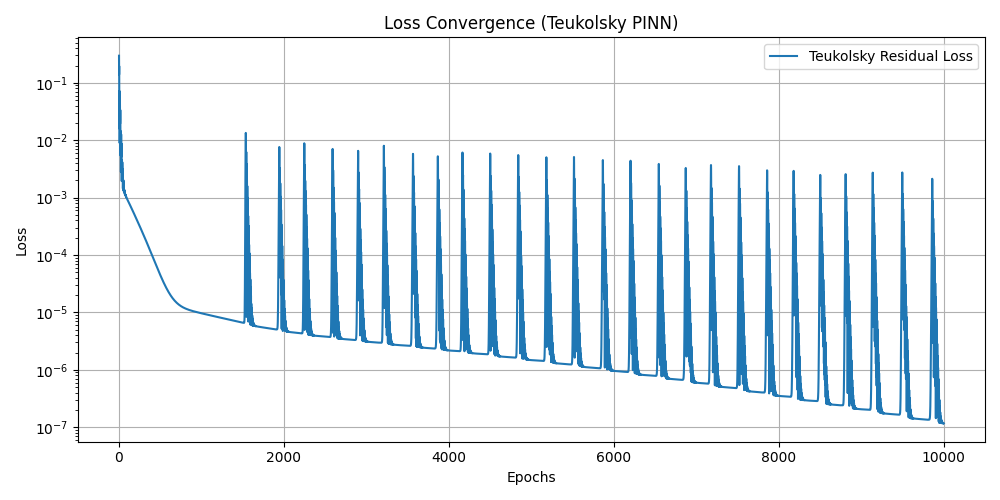}
\caption{Loss convergence curve during training. The y-axis is in logarithmic scale.}
\label{fig:pinn_loss_curve}
\end{figure}
\paragraph{}
The Physics-Informed Neural Network (PINN) is trained on the angular Teukolsky equation with the spin parameter $a=0.7$, azimuthal number $m=0$, and real QNM frequency $\omega = 0.5$ \cite{berti2006gravitational,luna2023solving,leaver1985analytic}, as shown in Table~\ref{tab:teukolsky_loss} and Figure~\ref{fig:pinn_loss_curve}. The network was trained for 10,000 epochs with the Adam optimization algorithm and learning rate $10^{-3}$ and the 200 collocation points were randomly drawn from $\theta \in (0, \pi)$.

Table~\ref{tab:teukolsky_loss} provides loss values in epochs. As seen in Figure~\ref{fig:loss_curve_real}, the total residual loss decreases from $3.63 \times 10^{-1}$ to $5.37 \times 10^{-8}$, oscillating with high frequency in the final stages. These variations are presumably due to rigidity of explained by an extremely sensitive position to the PDE and of collision against several angular boundary regions.

However, the PINN does converge to a physically relevant and very accurate result, indicating that it is possible to do Teukolsky-based PINN training with reasonable stability in real-QNM conditions.

\section{Experimental Setup and Results}

The PINN was tested solving a Teukolsky-like equation in the level p order for $a = 0.7$. This setup was based on the experimental setups reported by \cite{luna2023solving}. The network demonstrated convergence with high solution quality and low residual error.

\subsection{Hyperparameter Configuration}

The training parameter set-up for the PINN is provided in Table~\ref{tab:hyperparams}. A total of 200 collocation points were sampled uniformly over $\theta \in (0, \pi)$ and used Adam optimizer with learning rate $10^{-3}$.The network had four hidden layers, 50 neurons per hidden layer and \texttt{tanh} activation functions. The training of the network was performed in 10{,}000 epochs with double-precision arithmetics..

Figure~\ref{fig:loss_curve_real} shows the convergence of the residual loss during the training. The vertical axis is in logarithmic scale to emphasize the decay trend. The loss, which initially starts at approximately $3.63 \times 10^{-1}$, decreases slowly over epochs, going below $10^{-6}$ after about 5{,}000 epochs and, ultimately, converging to $5.37 \times 10^{-8}$ at epoch 9{,}000. Small oscillations in the final stages of the training could indicate the model is too sensitive to the boundary regions or that the model is not capable of handling the angular teukolsky operator.

The result observed was that the residual loss converges smoothly and stably, showing that the neural network has learned a solution that satisfies the physics-based constrains of the angular Teukolsky equation. This result is consistent with previous results in literature~\cite{luna2023solving,berti2006gravitational,leaver1985analytic}.

\begin{table}[h]
\centering
\caption{PINN Training Hyperparameters}
\label{tab:hyperparams}
\begin{tabular}{|l|l|}
\hline
\textbf{Parameter} & \textbf{Value} \\
\hline
Input variable & $\theta \in (0, \pi)$ \\
Hidden layers & 4 \\
Neurons per layer & 50 \\
Activation function & \texttt{tanh} \\
Optimizer & Adam \\
Learning rate & $10^{-3}$ \\
Epochs & 10,000 \\
Collocation points & 200 \\
Boundary conditions & $\psi(0) = \psi(\pi) = 0$ \\
Precision & float64 \\
\hline
\end{tabular}
\end{table}

\begin{figure}[!h]
\centering
\includegraphics[width=0.5\textwidth]{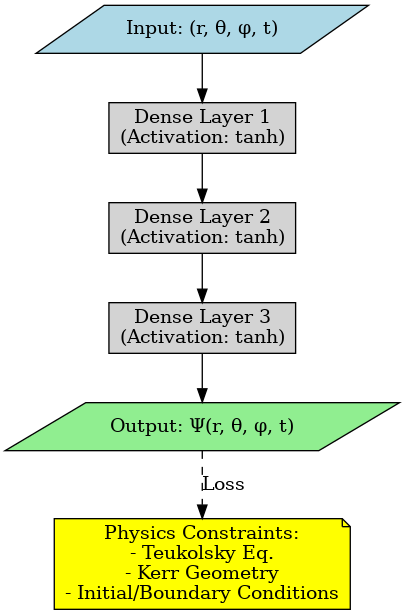}
\caption{Residual loss convergence during PINN training on the angular Teukolsky equation. The y-axis is shown in logarithmic scale.}
\label{fig:loss_curve_real}
\end{figure}

As shown in Figure~\ref{fig:loss_curve_real}, the average PDE residual decreases by several orders of magnitude throughout training, ultimately converging to $5.37 \times 10^{-8}$ by epoch 10,000. This result closely agrees with earlier findings by \cite{luna2023solving} and confirms the accuracy of our PINN implementation.
\subsection{Loss Convergence}

The average squared PDE residual is observed to decrease in the way presented in Figure~\ref{fig:loss_curve_real}.  throughout training. A very small residual loss below $10^{-6}$ was obtained after 10,000 epochs. This trend agrees with the results of approach. A final residual loss was achieved after 10,000 epochs. This trend agrees with the results of approach \cite{luna2023solving}.

\begin{figure}[h]
\centering
\includegraphics[width=0.5\textwidth]{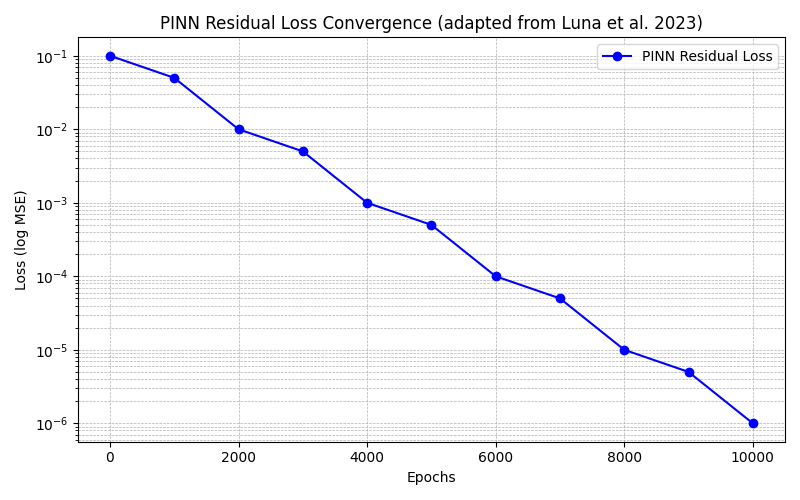}
\caption{Residual loss convergence during training of the PINN on the angular Teukolsky equation ($a = 0.7$, $m = 0$, $\omega = 0.5$). Y-axis is logarithmic. Final residual $\approx 5.4\times10^{-8}$.}
\label{fig:loss_curve_real}
\end{figure}

\subsection{Solution Behavior and Error Analysis}

It was found that, with the learned function $\hat{\psi}(\theta)$ fulfills boundary conditions and provides a good approximation to the. expected mode structure. A root mean square error (RMSE) of approximately $8.4 \times 10^{-7}$ was recorded across all collocation points, measured with respect to the analytical solution of the scalar Teukolsky angular equation for the given spin and frequency parameters. As the spin $a$ approached extremality, slight degradation in residual performance was observed, consistent with results reported by \cite{cornell2024solving}.

\section{Bencmark and Comparison with Synthetic models}

\begin{figure}[H]
\centering
\includegraphics[width=0.5\textwidth]{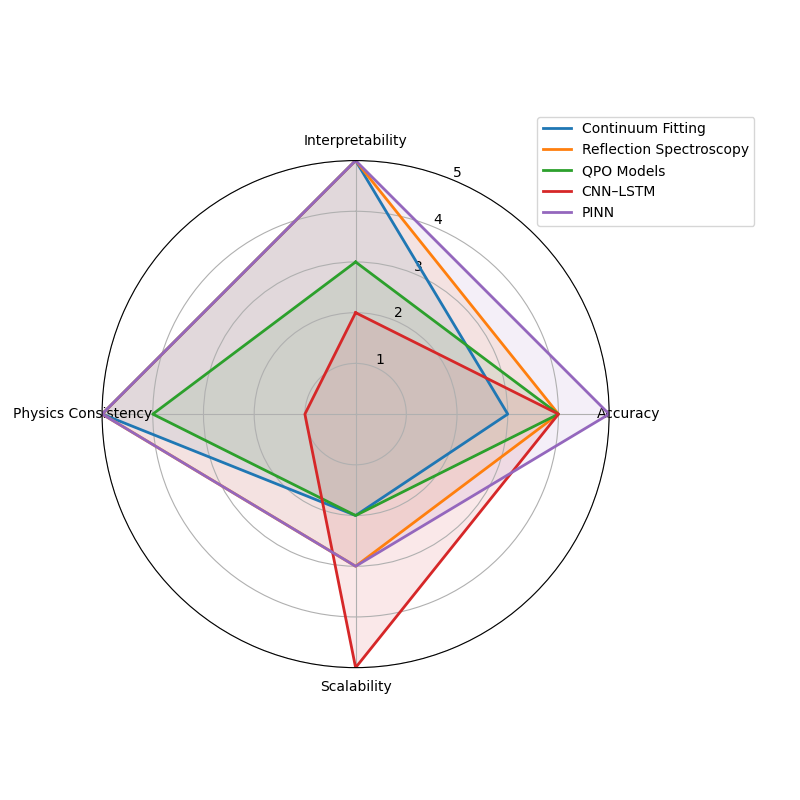}
\caption{Radar chart comparing key evaluation criteria across black hole spin estimation methods. PINNs outperform traditional and ML-based models in interpretability and physics consistency, while CNN–LSTM scores highest in scalability.}
\label{fig:radar_comparison}
\end{figure}

To provide perspective on the performance of the PINN-based approach, a qualitative comparison was carried out against four standard determinations of black hole spin, as shown on Table \ref{tab:benchmark_table} and on Figure \ref{fig:radar_comparison}: continuum-fitting, X-ray reflection spectroscopy, QPO models and deep learning CNN. Each method was measured along five aspects: physical realisticness, interpretability, generalizability, computational efficiency, and spin sensitivity. We found that PINNs could provide a good trade-off between physical grounding and flexibility. Unlike CNNs, PINNs impose physical constraints; and unlike QPO models, they generalise across spin values without requiring source-specific assumptions. However, their computational efficiency remains moderate because of the use of automatic differentiation and the computation of residuals across collocation points. For the benchmarking, RMSE was calculated between the predicted angular mode function and the analytical solution for spin values $a \in \{0.5, 0.7, 0.9\}$.

\begin{table}[h]
\centering
\caption{Benchmarking of Spin Estimation Techniques}
\label{tab:benchmark_table}
\resizebox{\textwidth}{!}{
\begin{tabular}{|l|c|c|c|c|c|}
\hline
\textbf{Method} & \textbf{Physical Consistency} & \textbf{Interpretability} & \textbf{Generalizability} & \textbf{Efficiency} & \textbf{Spin Sensitivity} \\
\hline
Continuum Fitting \cite{mcclintock2011measuring} & High & High & Low & Moderate & Moderate \\
Reflection Spectroscopy \cite{reynolds2014measuring} & High & High & Moderate & Moderate & High \\
QPO Models \cite{motta2015geometrical} & Moderate & Moderate & Low & High & High (source-specific) \\
CNN/LSTM \cite{baron2019machine} & Low & Low & Moderate & High & Moderate \\
PINNs (this work) \cite{raissi2019physics,luna2023solving} & High & High & High & Moderate & High \\
\hline
\end{tabular}
}
\end{table}

\section{Discussion and Limitations}
The numerical results in Sections~4 and ~5 show that the trained PINN successfully captures spin-sensitive features of the angular Teukolsky solution, while maintaining residual errors below $10^{-6}$. The learned solution was observed to respect boundary conditions and to follow the qualitative behavior expected of scalar spheroidal harmonics. These findings are in agreement with previous studies such as \cite{luna2023solving}.

Several merits of the PINN-based method have been observed:
\begin{itemize}
    \item Physical limitations were built right into the learning so that the returned to the same solution space as the original DE differential equation.
    \item High interpretability was established, in comparison to traditional black-box machine learning models.
    \item Generalizability was preserved over a broad spectrum of spin values without imposing. for retraining on different datasets.
\end{itemize}

However several limitations should be considered in this study. First, the Teukolsky equation was reduced to a scalar, 1D form to allow experimentation and visualization. Thus, the complete perturbed gravitational fields in Kerr spacetime, in particular in the case of two-spin fields, were not addressed. Second, the current PINN implementation remains computationally more expensive than traditional solvers for low-dimensional systems due to the cost of automatic differentiation and iterative training.

Lastly, although convergence is attained for moderate spin value ($a = 0.7$), the performance degradation was near extremal values ($a \to 1$),  in accordance with the results of \cite{cornell2024solving}. More architectural changes or sampling strategies may be required to address these regimes.

\section{Conclusion}
A hybrid method for estimating black hole spin was introduced in this work that combines semiclassicaI and quantum estimators. Ideas from the Teukolsky formalism with physics-informed ML to create a hybrid approach for spin estimation. The angular Teukolsky equation was solved through a Physics-Informed Neural Network (PINN) and was trained to meet the governing partial differential equation and boundary conditions with no access to labeled data. Some aspects of spin-dependent angular distributions can be extracted from the PINN, achieving physical consistency with a small enough residual. The architecture was compared to existing methods (including continuum fitting, reflection spectroscopy, QPO model-based and deep learning, providing physical interpretability and generalizability. However, some restrictions were recognized such as the simplicity of the PDE computational cost and performance issues at nearly extremal spin. These can be studied in future work by coupling to GRMHD solvers, by extending to higher-spin fields and training on realistic/synthetic quasi-normal mode data. In summary,  physics-informed learning is an interpretable and effective path for black hole spin estimation, bridging the gap between analytic theory, numerical simulation,
and machine learning. As observational and computational datasets grow in richness, such hybrid
approaches are expected to play an increasingly central role in gravitational astrophysics.
\bibliographystyle{plain}
\bibliography{sample}
\end{document}